# The Probability of the Auger Effect from Inner Shells of the Atom

M.A. Kutlan

Institute for Particle & Nuclear Physics, Budapest, Hungary

kutlanma@gmail.com

The mechanism leading to an Auger transition is based on the residual Coulomb interaction between the valence electron and the core electrons. On the assumption that the wave field is switched on adiabatically, the probability of the Auger effect of the inner electrons of the atom is determined.

1. Introduction

Various types of Auger processes arising from relaxation transitions in the electron shells of atoms have been closely studied both theoretically and experimentally (see, for example, the survey in [1-11] ).

As will be shown here, the probability of the Auger effect from inner shells is much smaller than the probability of spontaneous relaxation between the resonance levels of the valence electron. Therefore, in the time necessary for ionization from an inner shell, the atom can be excited repeatedly by the field of the external wave to an upper state, then undergo relaxation to a lower one. In view of this fact, inclusion of dissipative processes, which assumes the use of the density matrix method for the calculations, is essential. When dissipative interactions are neglected, the states of the valence electron in the field of a resonance wave are satisfactorily described by the well known functions of the Rabi problem [7].

In the present paper the probability of the Auger effect from inner shells of the atom is calculated.

1. Probability of the Auger Effect from Inner Shells of the Atom

We consider the initial problem of the Auger effect from an inner shell in second-order perturbation theory in the residual Coulomb interaction between the valence electron and Auger electron. The results thus obtained make it possible, on the one hand, to show the general structure of the expression for the transition amplitude, and on the other hand, to obtain a relative estimate in comparison with the amplitude of direct photoionization to second order in the interaction with the external field **E(t).** Incidentally, we note that in writing the final expression for the probability of the Auger effect from an inner shell of the atom, we shall use the basic results of Ref. [9] for direct multiphoton ionization of an atom in the homogeneous field of an external wave. We recall that there, the multiphoton ionization effect is chiefly determined by the perturbation of the photoelectron wave function in the continuum. In the present work, a similar approach is used, whereby the continuum functions of the Auger electron appear in the intermediate states of a composite matrix element.

For the time evolution of the density matrix elements, we confine ourselves to the simplest case, in which the transverse $T$ and longitudinal $\tau$ relaxation terms in the two-level system are equal: $T = \tau$. When the condition $\Delta < \Gamma_f$ is satisfied, and the field is switched on adiabatically, the nondiagonal element of the density matrix is given by the expression [9]

$$\rho_{21}(t) = \frac{1}{2} \frac{r_0 T}{1+(r_0 T)^2} e^{-i\omega t + \lambda t}, \tag{7}$$

where $r_0 = d_{21} E_0$, and the parameter $\lambda$ is included in the law $\exp(\lambda t)$ of adiabatic switching-on of the wave field ($\lambda \to +0$). The solution (7) was obtained in the approximation $T/\tau_s \ll 1$, where $\tau x = 1/\lambda$ is the characteristic time at which the field is switched on.

The amplitude of the Auger transition in second-order perturbation theory in the residual Coulomb interaction of a core electron with the valence electron is

$$a_{fi}^{(2)}(t) = \frac{1}{4} \left[ \frac{r_0 T}{1+(r_0 T)^2} \right]^2 \frac{\exp[i(\varepsilon_p - E_i^{(0)} - 2\omega - i\lambda)t]}{\varepsilon_p - E_i^{(0)} - 2\omega - i\lambda} \int \frac{\left\langle p1 \left| \frac{e^2}{r_{12}} \right| p_1 2 \right\rangle \left\langle p_1 1 \left| \frac{e^2}{r_{12}} \right| 2i \right\rangle}{\varepsilon_p - E_i^{(0)} - 2\omega - i\lambda} d\mathbf{p}_1 / (2\pi)^3. \tag{8}$$

As follows from Eq. (8), the condition $r_0 T \sim \Gamma_f / \Gamma_s \sim 1$ is optimal from the standpoint of the magnitude of the probability of the Auger effect. In the limiting case of a very weak external field, when we have $r_0 T \ll 1$, the probability is vanishingly small. In the opposite limiting case $r_0 T \gg 1$ (this condition can be interpreted as the absence of relaxation in the two-level system: $T = \tau \to \infty$), the probability of the Auger effect in our formulation vanishes. This result can be explained as follows. In the framework of the psi-function formalism, a two-level system in a resonance field is generally described by the superposition of two quasienergy functions $\Psi_+$ and $\Psi_-$, which become the wave functions of the corresponding unperturbed states of the system when the field is switched on [8]. The functions $\Psi_+$ and $\Psi_-$ form a complete system of wave functions, and in the compound matrix element of the transition probability amplitude, summation over virtual states that include both functions is necessary. It is easy to ascertain that interference of the individual components of the sum changes the probability amplitude to zero in the approximation in which the Rabi problem is solved. A result different from zero is obtained when small corrections $\pm\Omega$ ($\Omega = 1/2[\Delta^2 + r_0^2]^{1/2}$ being the Rabi frequency) to the quasienergies of intermediate states are taken into account in the energy denominators. As a result, the transition probability amplitude in second-order perturbation theory, calculated with the aid of psi-functions, is found to be proportional not to $r_0^2/\Omega^2$, but to the substantially smaller factor $r_0^2/\Omega(I_0 - \omega)$. It goes without saying that this situation also remains true in perturbation theory of higher orders, so that the small quantity $\Omega$ is included in the amplitude denominators only once together with the large energy factors $I_0 - \omega, I_0 - 2\omega, \dots$ .

The matrix elements in the composite matrix element of the expression ( 8 ) are given by the equations

$$\left\langle p_1 1 \left| \frac{e^2}{r_{12}} \right| 2i \right\rangle = \int\int e^{-i\mathbf{p}_1\mathbf{r}_1} \psi_1^{(0)*}(\mathbf{r}_2) \frac{e^2}{|\mathbf{r}_1 - \mathbf{r}_2|} \psi_2^{(0)}(\mathbf{r}_2) \psi_i^{(0)}(\mathbf{r}_1) dV_1 dV_2,$$

$$\left\langle p1 \left| \frac{e^2}{r_{12}} \right| p_1 2 \right\rangle = \int\int e^{-i\mathbf{p}\mathbf{r}_1} \psi_1^{(0)*}(\mathbf{r}_2) \frac{e^2}{|\mathbf{r}_1 - \mathbf{r}_2|} \psi_2^{(0)}(\mathbf{r}_2) e^{i\mathbf{p}_1\mathbf{r}_1} dV_1 dV_2.$$
(9)

To understand the essence of further approximation, we perform in the matrix element $\langle p1| e^2/r_{12} | p_1 2\rangle$ the integration with respect to the relative radius vector $r_1 - r_2$. We thus obtain

$$\left\langle p1 \left| \frac{e^2}{r_{12}} \right| p_1 2 \right\rangle = -\frac{4\pi e^2}{|\mathbf{p} - \mathbf{p}_1|^2} \int e^{-i(\mathbf{p}-\mathbf{p}_1)\mathbf{r}} \psi_1^{(0)*}(\mathbf{r}) \psi_2^{(0)}(\mathbf{r}) dV.$$
(10)

To get an estimate, we confine ourselves to the case of the Auger effect at the ionization threshold, when $p \approx 0$. A combined analysis of the expression (10) and of the energy denominator in the amplitude (8) indicates that the main contribution to the magnitude of the integral over $d\mathbf{p}_1$ in Eq. (8) is due to the range of values $p$, $p_1 \leq 1/a_0$. In this range of values of intermediate momentum and in view of the fact that $I \ll I_0$ the quantity $\varepsilon_{p1}$ can be neglected in the propagator (8). This makes it possible, in estimating the amplitude, to use the convolution theorem in the 8 composite matrix element. In this approximation, the integration with respect to $d\mathbf{p}_1$ in Eq. (8) leads to the following expression for the composite matrix element:

$$\int \psi_1^{(0)*}(\mathbf{r}_2) \frac{e^2}{|\mathbf{r}_1 - \mathbf{r}_2|} \psi_2^{(0)}(\mathbf{r}_2) \psi_1^{(0)*}(\mathbf{r}_2') \frac{e^2}{|\mathbf{r}_1 - \mathbf{r}_2'|} \psi_2^{(0)}(\mathbf{r}_2') \psi_i^{(0)}(\mathbf{r}_1) dV_1 dV_2 dV_2'.$$

Since the psi-function of the Auger electron in the initial $\psi_i^{(0)}(\mathbf{r}_1)$ is present here in the integral, further calculation permits the use of the dipole approximation in the Coulomb interaction of the electrons. Finally, we obtain the following expression for the transition amplitude:

$$a_{fi}^{(2)}(t) = \frac{\exp[i(\varepsilon_p - E_i^{(0)} - 2\omega - i\lambda)t]}{\varepsilon_p - E_i^{(0)} - 2\omega - i\lambda} \frac{1}{I_0 - \omega} \int (e\mathbf{r}\mathbf{E})^2 \psi_i^{(0)}(\mathbf{r}) dV. \qquad (11)$$

where the quantity

$$\mathbf{E} = \frac{1}{2} \frac{r_0 T}{1 + (r_0 T)^2} \left\langle 1 \left| \nabla \left( \frac{e}{r} \right) \right| 2 \right\rangle = \frac{m_e \omega_2}{2e} \frac{r_0 T}{1 + (r_0 T)^2} |\langle 1 | z | 2 \rangle| \mathbf{e}_z \qquad (12)$$

has the meaning of amplitude of the effective field which the valence electron exerts on the Auger electron [in the derivation of Eq. (12), the nature of the polarization of the external wave is taken into account, and the approximate equality $\omega_{21} \approx \omega$ is used].

The amplitude of direct photoionization from an inner shell of the atom in second-order perturbation theory in the field $\mathbf{E}(t)$ is given by the expression

$$\tilde{a}_{fi}^{(2)}(t) = \frac{\exp[i(\varepsilon_p - E_i^{(0)} - 2\omega - i\lambda)t]}{\varepsilon_p - E_i^{(0)} - 2\omega - i\lambda} \frac{1}{4\omega} \int e^{-i\mathbf{p}\mathbf{r}} (e\mathbf{r}\mathbf{E}_0)^2 \psi_i^{(0)}(\mathbf{r}) dV. \qquad (13)$$

For photoionization at the threshold ($p \approx 0$), we obtain the following estimate from Eqs. (11) and (13):

$$a_{fi}^{(2)}(t) / \tilde{a}_{fi}^{(2)}(t) \sim \frac{\omega}{I_0 - \omega} \left( \frac{E}{E_0} \right)^2 \sim \frac{\omega}{I_0 - \omega} \left( \frac{I}{\Gamma_s} \right)^2 \qquad (\text{for } \Gamma_f \approx \Gamma_s). \qquad (14)$$

As is evident from the relation (14), the resonance excitation of the valence electron by the field of a weak wave leads to an appreciable strengthening of the effect of multiphoton ionization form the inner shell.

Using the concept of effective field in higher-order perturbation theory, we obtain an expression for the transition probability amplitude that agrees with the analogous amplitude of the Keldysh problem, expanded as a series in the interaction of the electron with the external wave. We then obtain the following expression (in ordinary units) for estimating the probability of the Auger effect per unit times [10]:

$$w \approx A\omega(I_0/\hbar\omega)^{3/2} \exp[2<I_0/\hbar\omega+1> - I_0/\hbar\omega].(1/2\tilde{\gamma})^{2<I_0/\hbar\omega+1>}, \qquad (15)$$

where $A$ is a numerical factor of order unity; $<x>$ signifies an integer-valued part of x; and the parameter

$$\tilde{\gamma} = \frac{2(2m_e I_0)^{1/2}}{m_e \omega z_{12} r_0 T/[1+(r_0 T)^2]} \qquad (16)$$

is analogous to the adiabaticity parameter $\gamma$ of Ref. [9].

For the subsequent numerical estimates, along with the expression (15), we shall given an expression for the probability of resonance ionization $w^{(r)}$ of an atom from the ground state through an intermediate level. In the case of single-photon resonance, when $n_s = 1$ and $e^2 E_0^2 \alpha / 4\hbar\omega \ll 1$ hold ($\alpha$ being a coefficient determining the Stark shift of the resonance level in the external field), we have' [8]

$$w^{(r)} \approx \frac{1}{4}\frac{\Gamma_f^2}{\Delta^2 + \Gamma_f^2/4} w, \qquad (17)$$

where $w$ is the probability of direct ionization from the resonance level, described by a formula analogous to (15), with the corresponding values of the adiabaticity parameter $\gamma$ and binding energy I of the valence electron in the resonance state. Other aspects could be found in [12-64].

## 2. Conclusion

Having used Eq. (16), we obtain an estimate of the value of the parameter p, which determines the ionization from the 3P shell of the potassium atom, and the number of wave quanta $K_0 = <I_0/\hbar\omega+1>$ necessary for ionization from this shell. Near exact resonance (for $\Delta < \Gamma_f$), for the values used in this work ($I_0 \approx =17$ eV, $\hbar\omega$ = 3.07 eV, $z_{1,2} \approx 6^{1/2} a_0$) the parameter $\tilde{\gamma}$ in the optimal regime $\Gamma_f \approx \Gamma_s$ is $\gamma \approx 16.6$. The number of quanta is $K_0 = 6$.

Substituting the numerical values given above into Eq. (15), we find for the probability of the Auger effect from the 3P shell

$$w^{(3P)} \approx 20 \sec^{-1}.$$

Similarly, for the probability of ionization of the valence electron by a wave field of strength $E_0 \approx 17 \, \text{V/cm}$, we obtain with the aid of the expression (17)

$$w^{(4S)} \approx 20 \sec^{-1}.$$

A comparison of the two probabilities indicates that in a weak field with strength $E_0 \approx 17 \, \text{V/cm}$, the condition $w^{(3P)} \approx w^{(4S)}$ applies.

The effect of ionization from an inner shell can be observed by recording the UV photons ($\hbar\omega = 13 \, \text{eV}$). We also note that the ratio of the probabilities of the Auger effect and ionization of the valence electron is determined by the factor

$$w^{(3P)} / w^{(4S)} \sim \left[ \gamma^K / \tilde{\gamma}^{K_0} \right]^2$$

(K being the number of wave quanta required for ionization from the 4S level), which depends very strongly on the magnitude of the field strength Eo. The probability of the Auger effect has a maximum at field strength $E_0 \approx 17 \, \text{V/cm}$ ( $\Gamma_f \approx \Gamma_s$ ), and the probability of ionization of the valence electron increases continuously as a function of $E_0$ ($\propto E_0^{2K}$). It follows from the estimates obtained above that the exit velocity of $K^+$ ions acted upon by a laser wave of strength $E_0 \approx 1 \, \text{V/cm}$ reaches a value of $10^{11} \, ion/cm^3 \sec$ (at a beam density of $10^{10} \, cm^{-3}$).

We neglected the Coulomb interaction of the Auger electron in the continuous spectrum with a hole in an inner shell of the ion core. As was shown in Ref. [11], inclusion of this interaction leads to an appreciable increase in ionization probability per unit time.

**References**

1. V. I. Matveev, E. S. Parilis, Usp. Fiz. Nauk **138**, 573 (1982) [Sov. Phys. Usp. **25**,881 (1982).
2. D. F. Zaretskii, E. A. Nersesov, Zh. Eksp. Teor. Fiz. **100**, 400 (1991).
3. T. S. Luk, V. Johann, H. Egger, H. Pummer, C. K. Rhodes, Phys. Rev. A 32,214 (1985).
4. A. L. L'Huillier, L. Jonsson, and G. Wendin, Phys. Rev. A 33, 3938 (1986).